\documentclass[fleqn,12pt,twoside]{article}
\usepackage[headings]{espcrc1}

\readRCS
$Id: espcrc1.tex,v 1.2 2004/02/24 11:22:11 spepping Exp $
\usepackage{graphicx}
\usepackage[figuresright]{rotating}
\usepackage{graphicx}

\newcommand{\AmS}{{\protect\the\textfont2
  A\kern-.1667em\lower.5ex\hbox{M}\kern-.125emS}}
\newcommand{\bea}{\begin{eqnarray}}
\newcommand{\eea}{\end{eqnarray}}
\newcommand{\be}{\begin{equation}}
\newcommand{\ee}{\end{equation}}                                             
\newcommand{\Tr}{\mathrm{Tr}}

\title{Chromomagnetic stability of the CFL phase with a meson current}

\author{A. Gerhold\address[MCSD]{ Department of Physics,
        North Carolina State University, \\ 
        Raleigh, NC 27695}
        \thanks{Work supported in part by the US Department 
	 of Energy grant DE-FG02-03ER41260.}
        }

\begin{document}

\maketitle

\begin{abstract}
We show that the color-flavor locked (CFL) phase
of dense quark matter is unstable with respect to the formation 
of a Goldstone boson current in the vicinity of the point
$m_s^2=2\mu\Delta$. In this Goldstone boson current phase all 
components of the magnetic screening mass are real.
\end{abstract}

\section{INTRODUCTION}

It is well known that the groundstate of three flavor
quark matter at very high baryon density is the color-flavor-locked (CFL)
phase \cite{CFL}. 
At somewhat lower densities it is necessary to take into 
account a finite strange quark mass, which leads to gapless fermions
in the spectrum if $\mu_s>\Delta$ (with $\mu_s={m_s\over2\mu}$) \cite{Alford:2003fq}.
The problem is that these gapless
fermion modes cause instabilities in current-current correlation functions 
\cite{Huang:2004bg,Casalbuoni:2004tb}.
In the present contribution (which is based on \cite{Gerhold:2006dt}) we show 
that for $\mu_s\sim\Delta$ the instability is resolved by the formation of a
non-zero Goldstone boson current. Similar currents were considered previously in 
\cite{Son:2005qx,Kryjevski:2005qq}. For large $\mu_s$ (when the gaps become small)
it has been argued that
the instability is resolved by the formation of a LOFF state \cite{Casalbuoni:2005zp}.

In this work we ignore homogeneous 
kaon condensation \cite{Bedaque:2001je}, which simplifies the caluclation of the dispersion laws.

\section{FREE ENERGY}

 We consider an effective lagrangian that describes the interaction 
of gapped fermions with background gauge fields
\cite{Kryjevski:2005qq,Kryjevski:2004jw}
\bea
\label{l_eff}
{\cal L } &=& 
     \Tr\left(\chi_L^\dagger(iv\cdot\partial -\hat{\mu}^L-A_e Q)\chi_L\right)
   + \Tr\left(\chi_R^\dagger(iv\cdot\partial -\hat{\mu}^R-A_e Q)\chi_R\right)
 \nonumber \\
 & & \mbox{}
  -i \Tr \left(\chi_L^\dagger \chi_L X v\cdot(\partial-iA^T)X^\dagger\right)
  -i \Tr \left(\chi_R^\dagger \chi_R Y v\cdot(\partial-iA^T)Y^\dagger\right)
 \nonumber \\
 & & \mbox{}
 -\frac{1}{2}\sum_{a,b,i,j,k} 
  \Delta_k\left(\chi_{L}^{ai}\chi_{L}^{bj}
                \epsilon_{kab}\epsilon_{kij}
               -\chi_{R}^{ai}\chi_{R}^{bj}
                \epsilon_{kab}\epsilon_{kij} + h.c. \right).
\eea
Here, $\chi_{L,R}^{ai}$ are left/right handed fermions with color 
index $a$ and flavor index $i$, $A_\mu$ are $SU(3)_C$ color gauge 
fields, and $\hat{\mu}^L=MM^\dagger/(2\mu)$, $\hat{\mu}^R=M^\dagger
M/(2\mu)$ are effective chemical potentials induced by the quark 
mass matrix $M$. The matrix $Q=\mathrm{diag}({2\over3},-{1\over3},
-{1\over3})$ is the quark charge matrix and $A_e$ is an electro-static 
potential. The fields $X,Y$ determine the flavor orientation of the 
left and right handed gap terms and transform as $X\to LXC^T$, $Y\to 
RYC^T$ under $(L,R)\in SU(3)_L \times SU(3)_R$ and $C\in SU(3)_C$, and 
$\Delta_k$ $(k=1,2,3)$ are the CFL gap parameters.

We will assume that $X=Y=1$ which excludes the possibility of kaon 
condensation. This assumption significantly simplifies the calculation 
of the fermion dispersion relations and the current correlation 
functions. It is possible to suppress kaon condensation by including 
a large instanton induced interaction \cite{Schafer:2002ty},
which does not change our results qualitatively \cite{Gerhold:2006dt}.

We wish to study the possibility of forming 
a Goldstone boson current. Gauge invariance implies that the free 
energy only depends on the combinations $\vec{J}_L=X(\vec{\nabla}
-i\vec{A}^T)X^\dagger$ and $\vec{J}_R=Y(\vec{\nabla}-i\vec{A}^T)Y^\dagger$. We 
will restrict ourselves to diagonal currents $\vec{J}_{L,R}$. Within 
our approximations the free energies of the vector and axial-vector 
currents $\vec{J}_L=\pm\vec{J}_R$ are degenerate. We will consider 
the pure vector current $\vec{J}_L=\vec{J}_R=\vec{A}^T$ with
\be
\label{cur}
  \vec{A}^{T}={\textstyle{1\over2}}\vec{\jmath}\,
   \left(\lambda_3+{\textstyle{1\over\sqrt{3}}}\lambda_8 
  -{\textstyle{1\over\sqrt{6}}}\lambda_0\right),
\ee
where $\lambda_A$ $(A=1,\ldots,8)$ are the Gell-Mann matrices 
and $\lambda_0=\sqrt{2\over3}{\bf 1}$.
This ansatz has the feature that it does not shift the energy of electrically 
charged fermion modes. An ansatz of this type is favored 
if electric neutrality is enforced. The energetically preferred solution 
is indeed close to the ansatz equ.~(\ref{cur}) \cite{Gerhold:2006dt}.

The free energy is given by
\be
\label{om1}
  \Omega={1\over G}(\Delta_1^2+\Delta_2^2+\Delta_3^2)
   +{3\mu^2 \jmath^2\over8\pi^2}-{\mu^2\over4\pi^2}
  \int dp\int_{-1}^1 dt 
  \sum_{i=1}^9(|\epsilon_i|-|p|), 
\ee
where $p=\vec v\cdot\vec p-\mu$, $t=\cos\theta$, and $\epsilon_i$ are the quasiparticle energies
which are obtained from the effective Lagrangian equ. (\ref{l_eff}).
The constant $G$ fixes the magnitude of the gap 
in the chiral limit, which we denote with $\Delta(0)$.

The integral in equ.~(\ref{om1}) is quite complicated. We assume that 
the external fields and currents are small and expand in a set of 
parameters $b\in\{\tilde A_3/\Delta_1, \tilde A_8/\Delta_1, \jmath/\Delta_1, 
(\Delta_2-\Delta_1)/\Delta_1, (\Delta_3-\Delta_1)/\Delta_1, (\mu_s-\Delta_1)
/\Delta_1\}$, until numerical convergence is achieved.

\begin{figure}
\includegraphics[width=9cm]{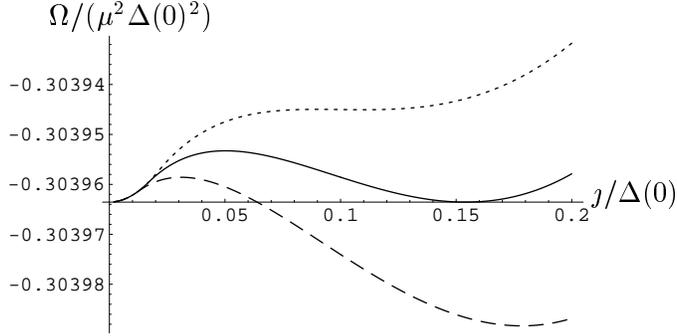}
\caption{$\Omega/(\mu^2\Delta(0)^2)$ 
as a function of $\jmath/\Delta(0)$ for $\mu_s=0.99\Delta(0)$ (dotted), 
$\mu_s=\mu_{s,crit}=0.9919\Delta(0)$ (continuous) and $\mu_s=0.994\Delta(0)$ (dashed).\label{figom}}
\end{figure}

We observe that $\Omega/(\mu^2\Delta(0)^2)$ is a function of $b/\Delta(0)$
and $\Delta_1/\Delta(0)$. This means that in order to study the effective 
potential as a function of dimensionless variables we do not have to 
specify the value of $\mu$ and $G$.
We solve the gap equations and neutrality conditions numerically.
Fig. \ref{figom} shows $\Omega(\jmath)$ for various values of $\mu_s$. 
We find that a nontrivial minimum appears for $\mu_s>\mu_{s,crit}=
0.9919\Delta(0)$.

\section{MEISSNER MASSES}

 In this section we study the stability of the current-current
correlation functions. The screening (Meissner) masses are given by
\bea
  (m^2_M)_{ab}^{ij} &=& {g^2\mu^2\over2\pi^2} \delta_{ab}\delta^{ij}
   + {g^2\mu^2 \over16\pi^2}\lim_{\vec k\to0}\lim_{k_0\to0}
   \int dp\int_{-1}^1 dt\int {dp_0\over{2\pi}}
       \nonumber \\[0.2cm]
  & &   \hspace{2cm}\mbox{} \times\mathrm{Tr} 
      \Big[ G^+(p) V_a^i G^+(p+k)V_b^j  
         +  G^-(p)\tilde V_a^i G^-(p+k)\tilde V_b^j \nonumber \\
  & & \hspace{2.7cm}\mbox{}
         + \Xi^+(p) V_a^i\Xi^-(p+k)\tilde V_b^j
         + \Xi^-(p)\tilde V_a^i \Xi^+(p+k) V_b^j\Big],
\eea
where $G^\pm$ and $\Xi^\pm$ denote ordinary and anomolous quark propagators
\cite{Gerhold:2006dt}, and $V_a^i$ and $\tilde V_a^i$ are quark-gluon vertices \cite{Gerhold:2006dt}.

\begin{figure}
\includegraphics[width=7.5cm]{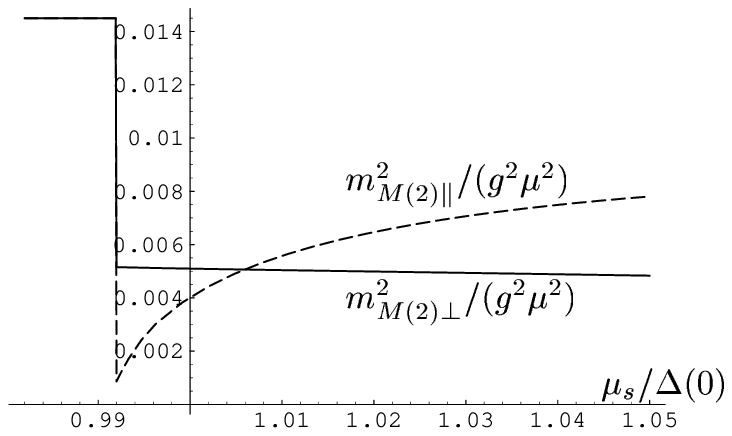}
\includegraphics[width=7.5cm]{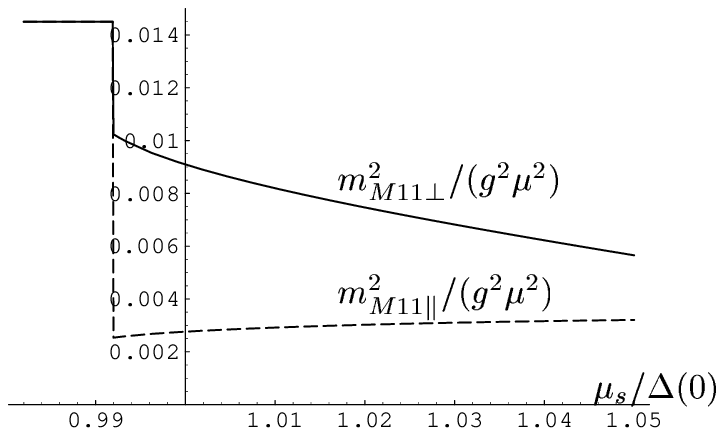}
\caption{Meissner masses squared $m_{M(2)\perp}^2/(g^2\mu^2)$,
$m_{M(2)\parallel}^2/(g^2\mu^2)$, $m_{M11\perp}^2/(g^2\mu^2)$ and
$m_{M11\parallel}^2/(g^2\mu^2)$ as functions of $\mu_s/\Delta(0)$.
\label{figmh}}
\end{figure}

For $\mu_s<\mu_{s,crit}$ one finds that the Meissner masses are real
\cite{Casalbuoni:2004tb,Gerhold:2006dt}.
In the presence of a finite current $\vec \jmath$ we may decompose the 
Meissner masses into a longitudinal and a transverse component,
\be
  (m_M^2)^{ij}=m_{M\perp}^2(\delta^{ij}-\hat \jmath^i\hat \jmath^j)
   +m_{M\parallel}^2\hat \jmath^i\hat \jmath^j. 
\ee
A chromomagnetic instability could occur for the Meissner masses with 
color indices 1,2,3 or 8 \cite{Casalbuoni:2004tb}.
Fig.~\ref{figmh} shows the ``dangerous'' components of 
the Meissner masses as functions of $\mu_s$, where $m^2_{M(2)}$ denotes
one of the eigenvalues in the 3-8-sector \cite{Gerhold:2006dt}.
We observe that all Meissner masses are real.

\section{CONCLUSIONS}

 We have shown that near the point at which gapless fermion modes 
appear in the spectrum the CFL phase becomes unstable with respect 
to the formation of a Goldstone boson current. We have computed 
the Meissner masses in the Goldstone current phase and found 
that all masses are real.
In this work we have not yet included the effect of a homogeneous kaon 
condensate, which is an important subject for future studies.

\section*{ACKNOWLEDGMENTS}
This work has been done in collaboration with T. Sch\"afer.

\end{document}